\documentclass{article}
\usepackage{spconf,amsmath,graphicx,amssymb}
\usepackage{multirow} 
\usepackage{url}
\usepackage{tabularx}
\usepackage{booktabs}
\usepackage{balance}
\usepackage{xcolor}
\usepackage{color}
\usepackage{flushend}
\usepackage{enumitem}
\usepackage{float,placeins}
\setlist{nosep, leftmargin=14pt}
\usepackage{mwe} 
\usepackage{hyperref}
\hypersetup{
    colorlinks=true,
    linkcolor=blue,
    filecolor=magenta,      
    urlcolor=cyan,
    pdftitle={Overleaf Example},
    pdfpagemode=FullScreen,
    }

\title{OCT2Confocal: 3D CycleGAN based Translation of Retinal OCT Images to Confocal Microscopy}
\name{Xin Tian, Nantheera Anantrasirichai, Lindsay Nicholson, Alin Achim \thanks{Xin Tian was sponsored by the China Scholarship Council.}}
\address{University of Bristol, United Kingdom}
\begin{document}
\ninept
\maketitle
\begin{abstract}
Optical coherence tomography (OCT) and confocal microscopy are pivotal in retinal imaging, each presenting unique benefits and limitations. In-vivo OCT offers rapid, non-invasive imaging but can be hampered by clarity issues and motion artifacts. Ex-vivo confocal microscopy provides high-resolution, cellular detailed color images but is invasive and poses ethical concerns and potential tissue damage. To bridge these modalities, we developed a 3D CycleGAN framework for unsupervised translation of in-vivo OCT to ex-vivo confocal microscopy images. Applied to our OCT2Confocal dataset, this framework effectively translates between 3D medical data domains, capturing vascular, textural, and cellular details with precision. This marks the first attempt to exploit the inherent 3D information of OCT and translate it into the rich, detailed color domain of confocal microscopy. Assessed through quantitative and qualitative evaluations, the 3D CycleGAN framework demonstrates commendable image fidelity and quality, outperforming existing methods despite the constraints of limited data. This non-invasive generation of retinal confocal images has the potential to further enhance diagnostic and monitoring capabilities in ophthalmology. Our source code and OCT2Confocal dataset are available at \url{https://github.com/xintian-99/OCT2Confocal}. 

\end{abstract}
\section{Introduction}
\label{sec:intro}
Optical coherence tomography (OCT) and confocal microscopy are key for three-dimensional retinal imaging, each revealing distinct aspects of retinal structure. OCT offers rapid, non-invasive imaging with detailed cross-sectional views and micrometer-scale resolution, widely used in ophthalmic studies. However, its clarity can be compromised by patient movement and other factors, impacting the precision needed for diagnosis and research \cite{retinareview2010}.

Confocal microscopy, available for both in-vivo superficial retinal imaging and ex-vivo, complements OCT with distinct advantages. The ex-vivo method, despite being more invasive due to tissue removal, produces vivid, colored images with superior resolution and no motion artifacts \cite{paula2010microstructure}. This approach is crucial for revealing deeper structural and cellular pathologies, and essential for in-depth ophthalmologic research and accurate pathology studies. However, it raises ethical concerns, carries risks of tissue damage, and can result in coloring inconsistencies during the staining process, complicating the interpretation of pathological features. These limitations, along with high costs, restrict the broader application of ex-vivo confocal microscopy.
    \begin{figure}[!htb]
    \centering
    \includegraphics[width=\columnwidth]{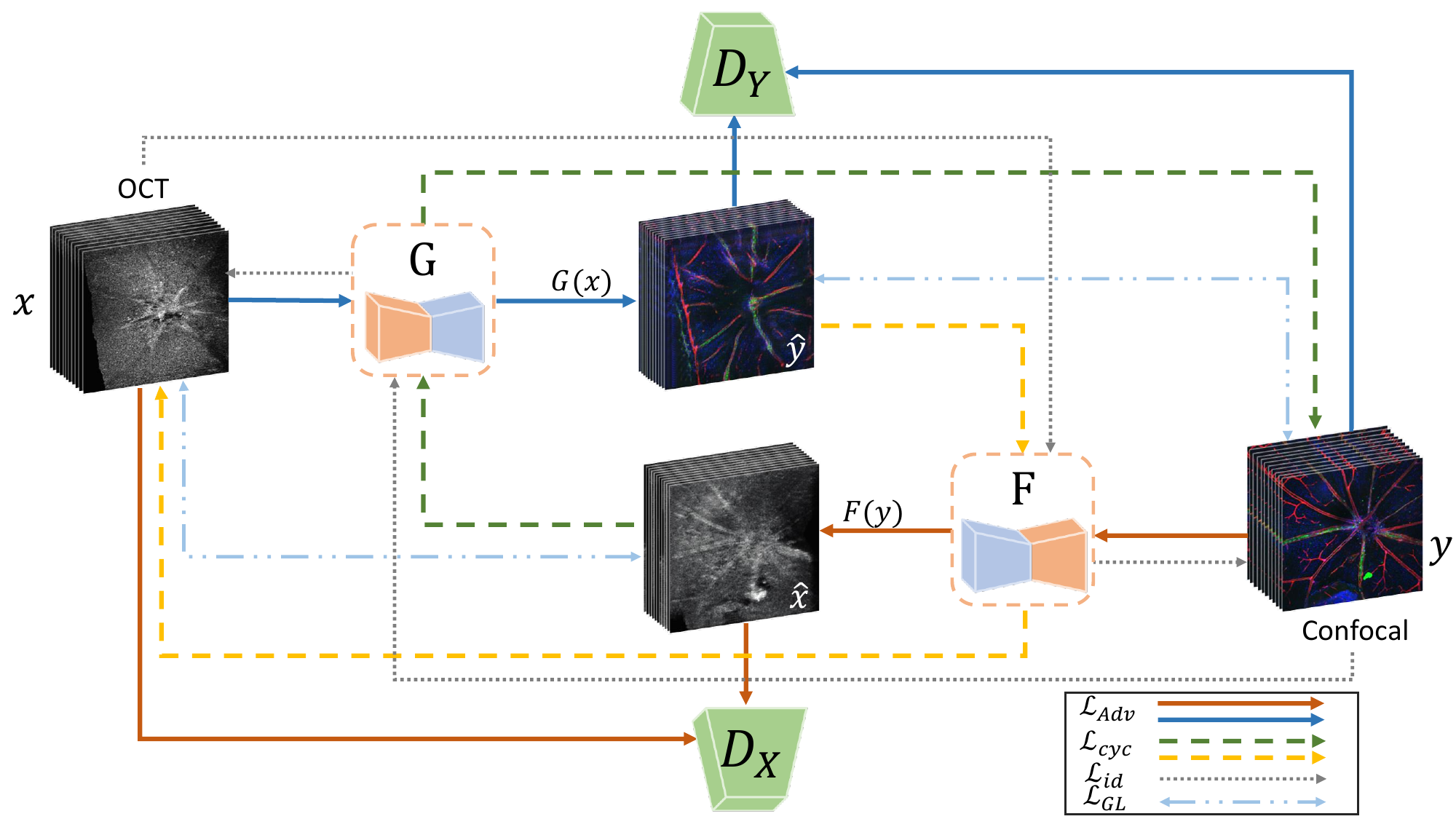}
    \vspace*{-0.5cm}
    \caption{The proposed 3D CycleGAN for OCT-to-Confocal.}
    \vspace*{-0.5cm}
    \label{fig:3D CycleGAN}
    \end{figure}
In response to the need for a swift and non-invasive method of obtaining high-resolution, detailed confocal images, we turn to the burgeoning field of deep learning-based medical image-to-image translation. This technology aims to transfer OCT images into the confocal domain, enriching them with high-resolution structural and cellular-level details. Common medical image-to-image translation approaches have evolved significantly with the advent of Generative Adversarial Networks (GAN) \cite{gan,shi2023translation}. For instance, the introduction of pix2pix\cite{pix2pix}, a supervised method based on conditional GANs, leveraging paired image as a condition for generating the synthetic image. However, obtaining such paired images can be challenging or even infeasible in many medical scenarios. Consequently, unpaired image-to-image translation methods, like CycleGAN \cite{cyclegan}, have emerged to fill this gap. While well applied in grayscale medical imaging modalities such as MRI and CT scans \cite{boulanger2021deep}, the translation between 3D volumetric grayscale and color (multichannel) images remains largely unexplored. 

In this work, we introduce a 3D CycleGAN approach to bridge the gap between in-vivo OCT and ex-vivo confocal volumetric retinal image translation. Utilizing our proposed OCT2Confocal dataset, which contains both OCT and corresponding confocal images, the model demonstrates its effectiveness in translating between 3D medical data domains. Our evaluations, both quantitative and qualitative, show that the model proficiently captures and translates vascular, textural, and cellular features from confocal images into the OCT domain. This work initiates a rapid and non-invasive method for generating retinal confocal images computationally, bypassing the need for traditional experimental procedures. This advancement holds significant potential for facilitating studies of immune responses and enhancing disease diagnosis in the retina.
\vspace*{-0.3cm}
\section{Proposed Methodology}

The proposed 3D CycleGAN method, an extension of the 2D CycleGAN architecture \cite{cyclegan}, employs 3D convolutions to simultaneously extract the spatial and depth information inherent in image stacks. The framework comprises two generators and two discriminators to map $X$ to $Y$ and vice versa. Given an OCT domain $X$ and a Confocal domain $Y$, the aim of our model is to extract statistic information from both $X$ and $Y$ and then learn a mapping $G: X \to\ Y$ such that the output $\hat{y}=G(x)$, where $x \in X$ and $y \in Y$. Another mapping $F$ transfers the estimated Confocal $\hat{y}$ back to the OCT domain $X$. The input images are processed as 3D stacks, and all learnable kernels within the network are three-dimensional, as depicted in Figure \ref{fig:3D CycleGAN}.

\subsection{3D CycleGAN Architecture}
\vspace*{-0.1cm}

\begin{enumerate}
\item [1)] \textbf{Generator}
The generator \( G \) employs a Convolution-InstanceNorm-ReLU layer, three downsampling layers, and 9 residual blocks (ResNet 9), starting with OCT input of 512×512×9×1 (grayscale) in this paper. Upsampling is achieved through three fractional-strided convolution layers, and a final convolution layer converts the output to multichannel (3 channels). Reflection padding is used to reduce edge artifacts. Experiment results confirmed that this configuration(ResNet 9) outperforms alternatives like U-Net\cite{unet} and WGAN-GP\cite{wgan} architectures. Generator \( F \) mirrors \( G \)'s design but outputs single-channel images for OCT reconstruction from 512×512×9×3 confocal inputs.


\item [2)] \textbf{Discriminator}
The discriminator networks in our model are adaptations of the 2D PatchGAN \cite{pix2pix}. In our implementation, the 3D PatchGAN assesses 70x70x9 voxel cubes from the 3D images to evaluate their authenticity.
\end{enumerate}

\vspace*{-0.2cm} \subsection{The Loss Function}

The total loss function is the weighted sum of the following four terms: i) adversarial loss \cite{gan} based on the Binary Cross-Entropy (BCE) loss, ii) cycle consistency loss \cite{cyclegan} with $L_1$ distance, iii) identity loss to maintain the consistency of an image when transformed within its original domain iv) gradient loss \cite{glloss} to enhance the textural and edge sharpness in the translated images.

\vspace*{-0.2cm}
\section{OCT2Confocal Dataset}
\begin{figure}[htb]
\includegraphics[width=\columnwidth]{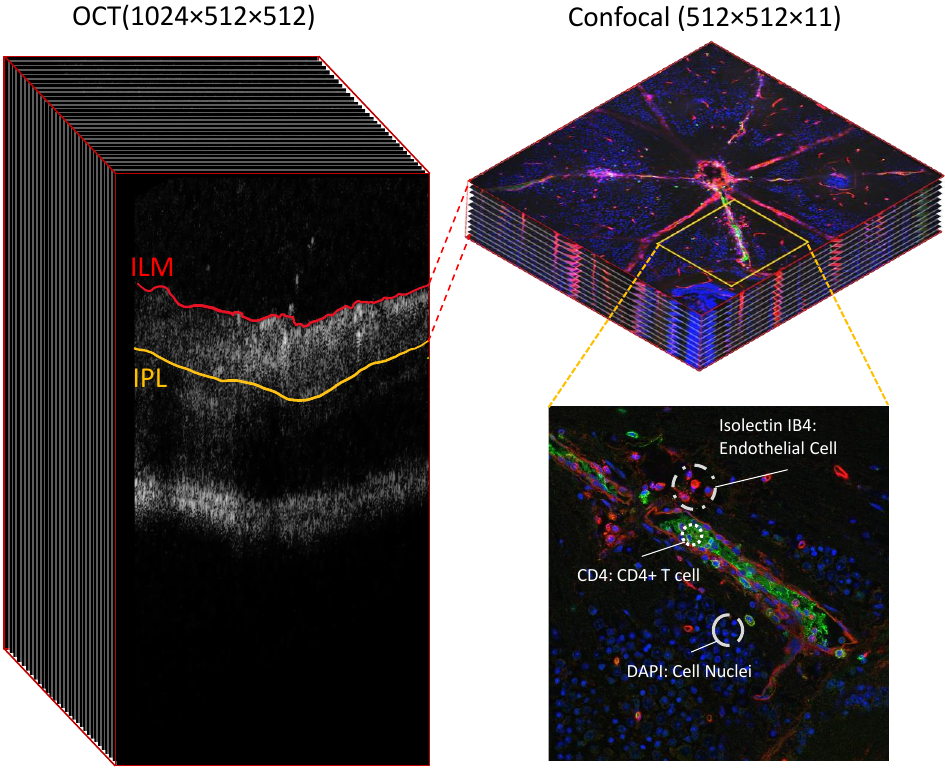}
\vspace*{-0.6cm}
\caption{OCT and Corresponding Confocal Image of Mouse A2R.}
\label{fig:A2R}
\vspace*{-0.5cm}
\end{figure}

\vspace*{-0.2cm}
We introduce the OCT2Confocal dataset, to the best of our knowledge, the first to include in-vivo grayscale OCT and corresponding ex-vivo colored confocal images. Images are acquired from C57BL/6 mice (A2L, A2R, B3R) with autoimmune uveitis. The OCT images, with a resolution of 512$\times$512$\times$1024 (\( H \times W \times D \)) pixels, were obtained using a Micron IV fundus camera featuring an OCT scan head and mouse-specific objective lens from Phoenix Technologies, California. These images are centered around the optic disc and characterized by in-vivo imaging artifacts. For image-to-image translation, the OCT images on day 24 between the retinal layers Inner Limiting Membrane (ILM) and Inner Plexiform Layer (IPL) are cropped to match the depth of confocal microscopy images. The confocal images, obtained post-euthanasia using a Leica SP5-AOBS microscope, feature staining for cell nuclei (DAPI - blue), CD4+ T cells (green), endothelial cells (Isolectin IB4 - red). The multichannel confocal images, providing comprehensive colored cellular detail, with their corresponding OCT images forms the training set, example illustrated in Figure \ref{fig:A2R}. These confocal images include resolutions of A2L at 512×512×14 pixels, A2R at 512×512×11 pixels, and B3R at 512×512×14 pixels, are captured between the ILM and IPL layers. Since captured in-vivo and ex-vivo, the OCT and confocal images are unpaired. Additionally, 22 OCT images without confocal images, also induced autoimmune uveitis, were used to assess the model's translation performance. This dataset initiates the application of OCT2Confocal image translation and holds the potential to deepen retinal analysis, thus improving diagnostic accuracy and monitoring efficacy in ophthalmology.

\section{Experiments Setup}
\vspace*{-0.2cm}
\subsection{Evaluation Methods}

\begin{itemize}
    \item [1)] \textbf{Quantitative Evaluation}
    Distribution-Based (DB) metrics, the Fréchet Inception Distance (FID) \cite{fid} and Kernel Inception Distance (KID) \cite{kid}, are utilized for assessing perceptual quality, with lower scores signifying higher image fidelity. Metrics like MAE, PSNR, and SSIM are unsuitable as they require voxel-wise paired comparisons. Specifically 768-dimensional (FID768) and 2048-dimensional (FID2048) vectors, were used.
    
    \item [2)] \textbf{Qualitative Evaluation}
     The qualitative assessment used a subjective test with 10 experts (5 ophthalmologists and 5 medical image processing experts). They ranked images from UNSB\cite{unsb}, 2D CycleGAN\cite{cyclegan}, and three variants of the proposed approach across 13 sets, focusing on authenticity, color code preservation, aesthetics, and artifact exclusion. The evaluation started with reviewing three image sets, each comprising an original OCT, a real confocal image, and five translated images. The next part is to rank 10 blind sets without real confocal references. The outcomes were quantified into a Mean Opinion Score (MOS) ranging from 0 to 100, where higher scores indicate more authentic and higher-quality translations.
    
\end{itemize}
\vspace*{-0.3cm} \subsection{Implementation Details}
\vspace*{-0.1cm}
The implementation was conducted in Python with the PyTorch library. Training and evaluation took place on the BlueCrystal Phase 4 supercomputer at the University of Bristol, featuring Nvidia P100 GPUs with 16 GB RAM, and a local workstation outfitted with RTX 3090 GPUs. Optimization utilized the Adam optimizer with a batch size of 1, and a momentum term of 0.5. Learning rates were set at \(2 \times 10^{-5}\), with an input depth of 9 slices and instance normalization was employed. Data augmentation includes horizontal flipping, random zooming (0.9-1.1 scale), and random cropping (512×512 from 522×522 pixels).

\begin{figure*}[htb]
\includegraphics[width=\textwidth]{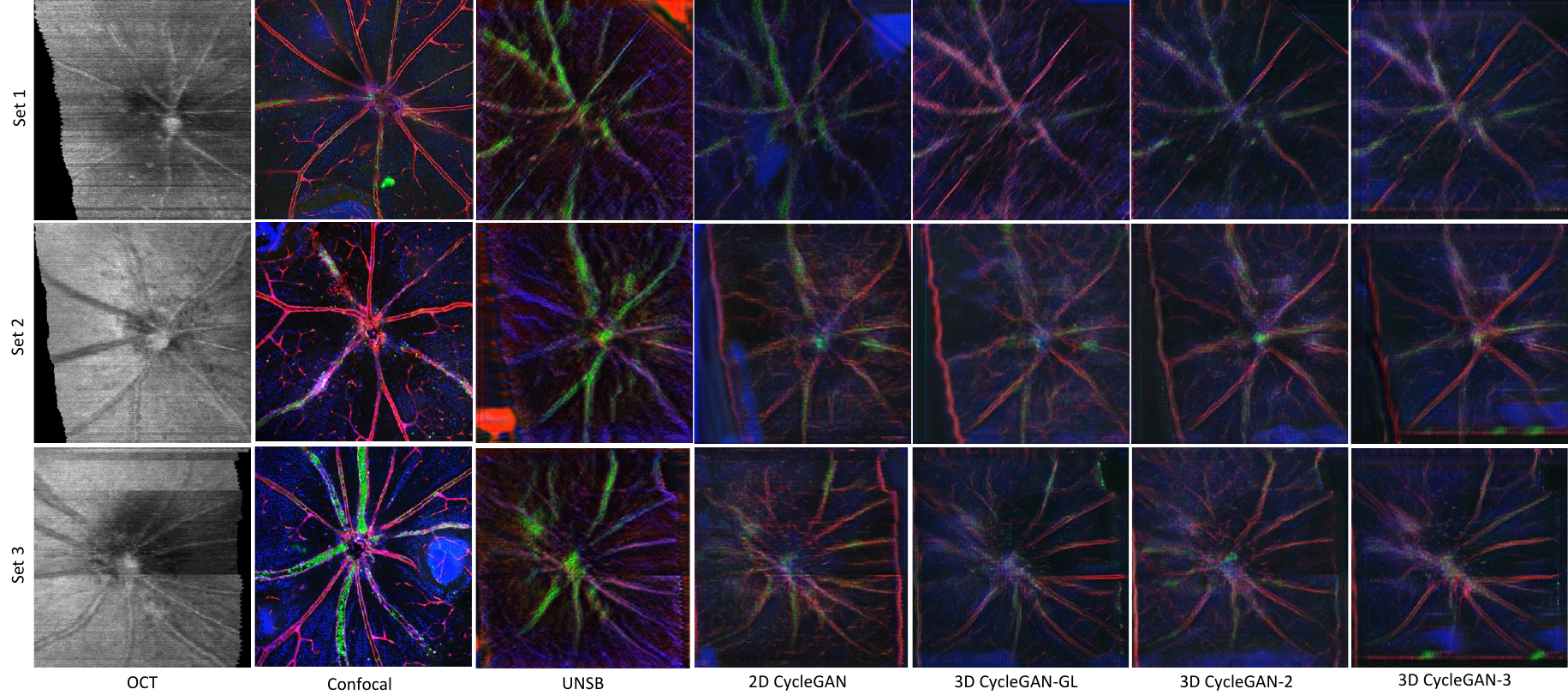}
\vspace*{-0.8cm}
\caption{Comparative Translation Results with Reference}
\label{fig:wref}
\vspace*{-0.5cm}
\end{figure*}

\begin{figure}[htb]
\includegraphics[width=\columnwidth]{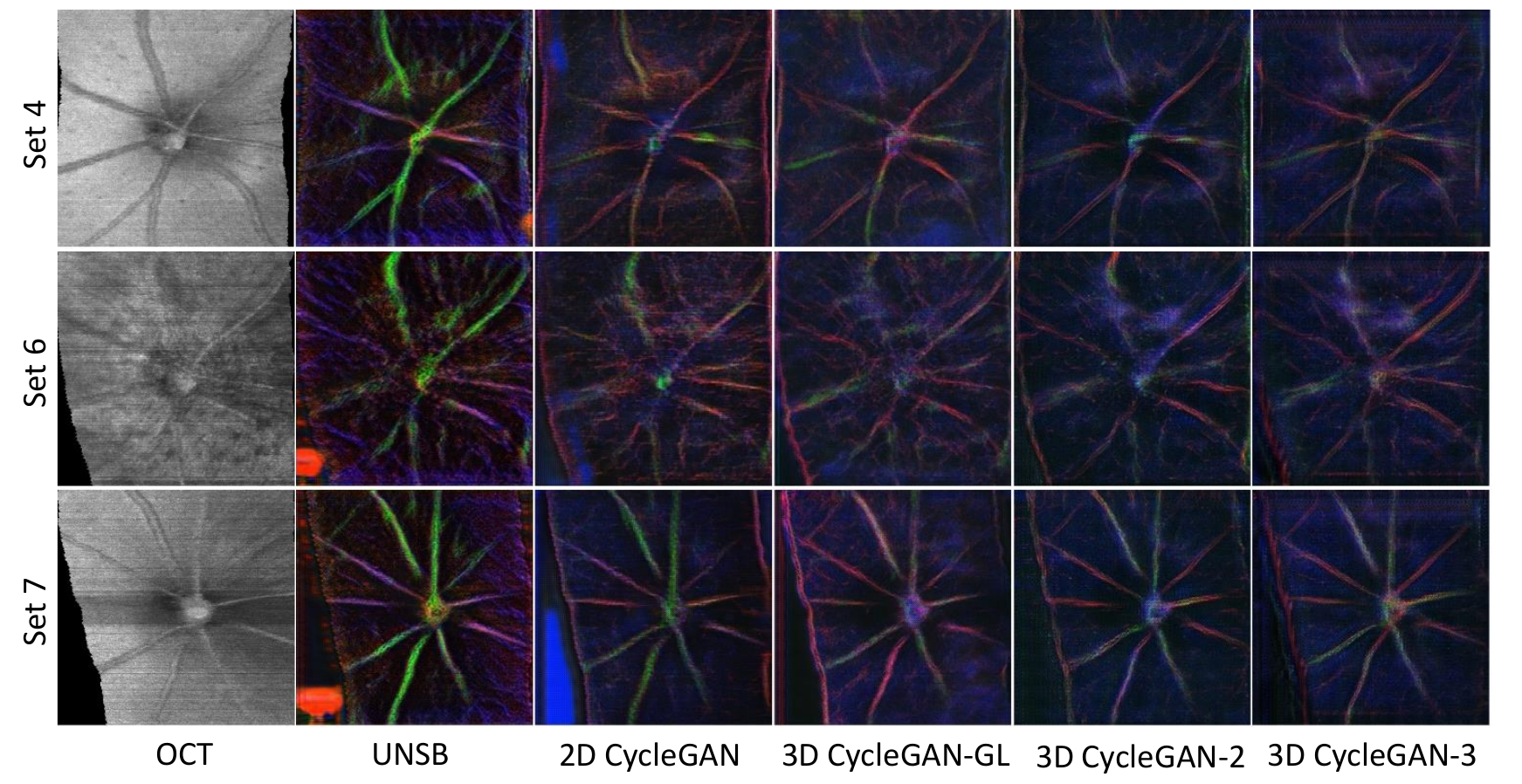}
\vspace*{-0.6cm}
\caption{Comparative Translation Results without Reference.}
\label{fig:woref}
\vspace*{-0.4cm}
\end{figure}
\vspace*{-0.4cm}\section{Results and Analysis}

We compared our method against the UNSB diffusion model \cite{unsb} and the conventional 2D CycleGAN \cite{cyclegan}, underscoring the effectiveness of the 3D network. We evaluated against 3D CycleGAN variants: 3D CycleGAN-2 with two downsampling layers without gradient loss, and 3D CycleGAN-GL with the same layers but including Gradient Loss. Our final model, 3D CycleGAN-3, is with three downsampling layers and gradient loss. Each model was retrained on the same dataset and configurations for consistency. For clearer visualization, results are displayed as fundus-like 2D projections from the translated 3D volume.
\vspace*{-0.2cm} \subsection{Ablation Study}
\begin{figure}[htb]
\includegraphics[width=\columnwidth]{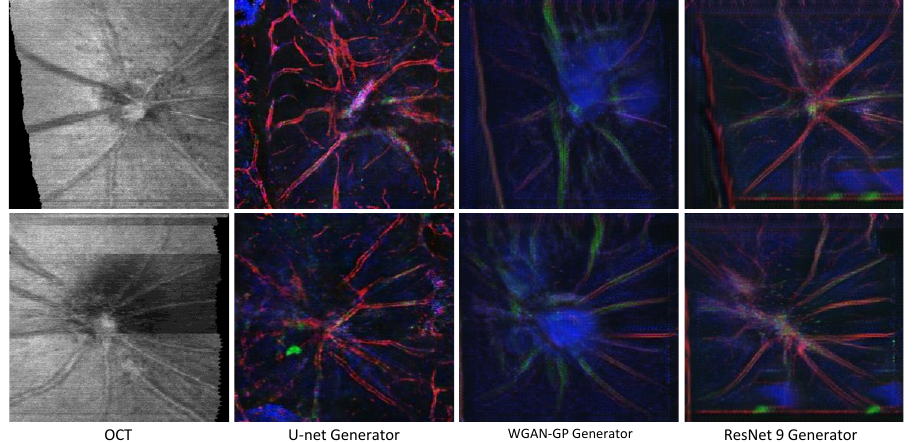}
\vspace*{-0.5cm}
\caption{Examples of translated results from different generators.}
\label{fig:ablation}
\vspace*{-0.5cm}
\end{figure}

\vspace*{-0.1cm}
In our experiments, the structure of the generator was found to have the most significant impact on the generated results, overshadowing other factors such as the hyperparameters of gradient loss and identity loss. Therefore, the ablation study in Table \ref{tab:ablation} focuses on the impact of generator architectures on the quality of synthesized images within the 3D CycleGAN framework. 
Among the U-Net, WGAN-GP, and ResNet 9 architectures, ResNet 9 demonstrates a distinct advantage in producing higher-quality images, as evidenced by its lower FID scores. While WGAN-GP attains the lowest KID score, visual assessment in Figure \ref{fig:ablation} shows that it still produces significant artifacts, underscoring the limitation of FID and KID metrics to fully assess image quality in medical imaging contexts. Detailed visual comparisons in Figure \ref{fig:ablation} reveal that U-Net architecture struggles with maintaining the clarity and shape of blood vessels, while the WGAN-GP does not consistently preserve color code, also introducing noticeable artifacts.

\begin{table}[htb]
\vspace*{-0.2cm}
\centering
\begin{tabular}{@{}cccc@{}}
\toprule
Generator & FID768 ↓ & FID2048 ↓ & KID ↓ \\ \midrule
U-Net & 1.135 & 178.445 & 0.182 \\
WGAN-GP & 1.202 & 173.142 & {\color[HTML]{FE0000}0.129} \\
ResNet 9 & {\color[HTML]{FE0000} 0.785} & {\color[HTML]{FE0000} 151.302} &  0.143 \\ \bottomrule
\end{tabular}%
\vspace*{-0.3cm}
\caption{Comparative Results of Different Generators Architecture in 3D CycleGAN. The best result is colored in {\color[HTML]{FF0000} red}.}
\label{tab:ablation}
\end{table}
\vspace*{-0.3cm}


\begin{table*}[htb!]
\centering
\label{allresult}
\resizebox{\textwidth}{!}{%
\begin{tabular}{@{}lcccccccccccc@{}}
\toprule
 & \multicolumn{4}{c}{W Ref} & \multicolumn{4}{c}{W/O Ref} & \multicolumn{4}{c}{Total} \\ \cmidrule(lr){2-5} \cmidrule(lr){6-9} \cmidrule(lr){10-13}  
\multirow{-2}{*}{Method} & FID768 ↓ & FID2048 ↓ & KID ↓ & MOS ↑ & FID768 ↓ & FID2048 ↓ & KID ↓ & MOS ↑ & FID768 ↓ & FID2048 ↓ & KID ↓ & MOS ↑ \\ \midrule
UNSB & 1.659 & 313.189 & 0.597 & 29.300 & 1.611 & 301.666 & 0.655 & 25.360 & 1.622 & 304.325 & 0.641 & 26.269 \\
2D CycleGAN & 1.547 & 225.302 & 0.300 & 36.667 & 1.420 & 231.048 & 0.326 & 41.630 & 1.449 & 229.722 & 0.320 & 40.485 \\
3D CycleGAN-GL & 1.281 & 202.795 & 0.267 & 50.400 & 1.170 & 169.556 & 0.215 & 49.550 & 1.195 & 177.227 & 0.227 & 49.746 \\
3D CycleGAN-2 & {\color[HTML]{0922DB} 0.852} & {\color[HTML]{FF0000} 149.486} & {\color[HTML]{FF0000} 0.144} & {\color[HTML]{0922DB} 53.967} & {\color[HTML]{0922DB} 0.890} & {\color[HTML]{0922DB} 166.473} & {\color[HTML]{0922DB} 0.160} & {\color[HTML]{0922DB} 52.860} & {\color[HTML]{0922DB} 0.881} & {\color[HTML]{0922DB} 162.553} & {\color[HTML]{0922DB} 0.156} & {\color[HTML]{0922DB} 53.115} \\
3D CycleGAN-3 & {\color[HTML]{FF0000} 0.766} & {\color[HTML]{0922DB} 154.756} & {\color[HTML]{0922DB} 0.153} & {\color[HTML]{FF0000} 56.867} & {\color[HTML]{FF0000} 0.780} & {\color[HTML]{FF0000} 155.188} & {\color[HTML]{FF0000} 0.156} & {\color[HTML]{FF0000} 56.350} & {\color[HTML]{FF0000} 0.777} & {\color[HTML]{FF0000} 155.089} & {\color[HTML]{FF0000} 0.155} & {\color[HTML]{FF0000} 56.469}\\
\bottomrule
\end{tabular}%
}
\vspace*{-0.3cm}
\caption{The performance of models evaluated by DB metrics FID scores at 768 and 2048 feature dimensions and KID scores, alongside the subjective MOS rating. The results are referred to categories with reference (W Ref), without reference (W/O Ref), and total image sets. The best result is colored in {\color[HTML]{FF0000} red} and the second best is coloured in {\color[HTML]{0922DB} blue}.}
\vspace*{-0.2cm}
\label{tab:result}
\end{table*}

\vspace*{-0.2cm} \subsection{Quantitative Evaluation}
\vspace*{-0.1cm}
From Table \ref{tab:result}, across all DB metrics, the 3D CycleGAN-3 model outperformed other methods, achieving the lowest FID and KID scores in all scenarios (With Reference, Without Reference, and total dataset). These results suggest that this model is relatively the most effective in aligning the statistical distribution of generated confocal images with those of real images, indicating higher image fidelity and better perceptual quality. The 3D CycleGAN-2 model follows as the second best, performing notably well in the with-reference scenario. Overall, the 3D CycleGAN models outperform the UNSB diffusion model and the 2D CycleGAN, demonstrating their enhanced ability to translate from in-vivo OCT to the ex-vivo confocal domain.

\vspace*{-0.2cm} \subsection{Qualitative Evaluation}
\vspace*{-0.1cm}
According to MOS in Table \ref{tab:result}, the 3D CycleGAN-3 model scored the highest across all three scenarios, as determined by the expert panel's rankings. This reflects the model's superior performance in terms of authenticity, detail preservation, and overall aesthetic quality. 

According to feedback from ophthalmologists and visual observations from Figure \ref{fig:wref} and Figure \ref{fig:woref}, all 3D CycleGAN models effectively preserve blood vessel clarity, shape, and color code more than other models. The 3D CycleGAN-3 model is reported to reflect the capacity for retaining more background detail and overall authenticity. In contrast, the 2D CycleGAN and UNSB models sometimes introduce random colors, disregard edges, and fail to adhere to the retinal vessels' color code. Notably, the 3D CycleGAN-3 model excelled in translating lower quality in-vivo OCT images, as seen in Set 6 of Figure \ref{fig:woref}, outperforming 2D and UNSB models. This showcases our model's ability to capture the deeper distributional relationships between OCT and confocal domains. It also minimizes the presence of artifacts which is critical for the utility of translated images in clinical settings.

\vspace*{-0.3cm}
\section{Conclusions and Future Work}

In this paper, we present the 3D CycleGAN framework as an effective tool for translating from the OCT to the confocal domain, effectively bridging in-vivo and ex-vivo imaging modalities. Though limited by dataset size, our quantitative and qualitative experiments show promising results in preserving essential image details, including blood vessel clarity and color code. We also introduce the OCT2Confocal dataset. Future efforts will focus on expanding the dataset for higher resolution outputs and optimizing the 3D framework for computational efficiency, aiming to advance early disease detection and diagnostics. Additionally, we plan to adapt the model for human OCT to confocal translation and apply the translated results to early disease detection and enhanced diagnostic practices.
\section{COMPLIANCE WITH ETHICAL STANDARDS}
\vspace*{-0.3cm}
All mice experiments were approved by the local Animal Welfare and Ethical Review Board (Bristol AWERB), and were conducted under a Home Office Project Licence.

\FloatBarrier
\bibliographystyle{IEEEbib}
\bibliography{refs}

\end{document}